\documentclass[review]{elsarticle}
\graphicspath{ {./figures/} }
\usepackage{hyperref}
\usepackage{float}
\usepackage{verbatim} 
\usepackage{apalike}
\restylefloat{figure}
\restylefloat{table}

\journal{Natural Language Processing Journal}

\biboptions{authoryear}

\begin{document}
\begin{frontmatter}

\begin{titlepage}
\begin{center}
\vspace*{1cm}

\textbf{BRAIN L: A book recommender system}

\vspace{1.5cm}

Jessie Caridad Martín Sujo (jessiecaridad.martin@salle.url.edu), Elisabet Golobardes i Ribé (elisabet.golobardes@salle.url.edu) \\

\hspace{10pt}

\begin{flushleft}
\small  
\begin{center} Research group on Smart Society \\ La Salle-Universitat Ramon Llull\\
Carrer de Sant Joan de la Salle, 42, Barcelona 08022, Spain \end{center}

\vspace{1cm}
\textbf{Corresponding Author:} \\
Jessie Caridad Martín Sujo\\
La Salle-Universitat Ramon Llull\\
Carrer de Sant Joan de la Salle, 42, Barcelona 08022, Spain \\
Email: jessiecaridad.martin@salle.url.edu

\end{flushleft}        
\end{center}
\end{titlepage}

\title{BRAIN L: A book recommender system}

\author{Jessie Caridad Martín Sujo\corref{cor1}}
\ead{jessiecaridad.martin@salle.url.edu}

\author{Elisabet Golobardes i Ribé}
\ead{elisabetgolobardes@salle.url.edu}

\cortext[cor1]{Corresponding author.}
\address{Research group on Smart Society \\La Salle-Universitat Ramon Llull, Barcelona 08022, Spain}

\begin{abstract}
Book sales in Spain have fallen progressively, which requires urgent changes to optimize the sales process as much as possible. This research proposes a new system, called Base of Reasoning in Artificial Intelligence with Natural Language (BRAIN L) focused exclusively on the publishing industry. The new field of knowledge of Artificial Intelligence (AI), Natural Language Processing (NLP), tecnología del Machine Learning is combined with Case-Based Reasoning (CBR) techniques for book recommendations. A model is developed to retrieve similar cases/books supported by NLP techniques for decision making. In addition, policies are implemented to keep the model evaluated by expert reviews, where the system not only learns with new cases, but these cases are real.
\end{abstract}

\begin{keyword}
Artificial Intelligence \sep Machine Learning \sep Case-Based Reasoning \sep Case-Based Recommendation \sep Natural Language Processing \sep Book Recommender
\end{keyword}

\end{frontmatter}

\section{Introduction}
\label{introduction}

In recent years, in Spain, reading time and the practice of it, as leisure for the population, has decreased significantly. A study by \cite{megias2018jovenes}, carried out jointly with the Reina Sofía Center for Adolescence and Youth and Fundación MAPFRE, gathers the opinion of 1,401 young people between the ages of 14 and 24 and different work groups. It shows that practically half of the young people affirm that, due to the use of social networks, study and reading time decreases. Even recently, systems have been created to analyze the time spent by readers, such as the one implemented in the study by \cite{martin2022correction}, where the authors show an 88\% effectiveness in predicting the permanence of readers in digital newspapers. In that study the data indicates that the larger content (length of the text) of the news along with an intermediate number of audios inserted in it, the greater the probability that a reader will be attracted to it. Obviously a book does not have the same characteristics as digital news and audio cannot be inserted. But these studies inspire us to look for solutions to once again captivate the reader in the literary world.

But why don't people read nowadays? This question may be based on multiple well-known factors such as prices, piracy, the impact of crises, among others. But, an essential point in this problem is found in the poor connection of the reader with the literary work.

The situation described above leads to the question: Would designing a book recommender based on reader behavior encourage more reading?

The main objectives of this research are:
\begin{itemize}
    \item Create a memory of cases, not only with the books to recommend but also the determination of a psychological profile of the reader.
    \item Evaluate the similarity of the text entered in the new CBR case with the texts of the case memory.
    \item Implement policies so that real cases are saved during the memory storage phase.
\end{itemize}

Starting from the basis of the research question and the stated objectives, it is hypothesized: It is possible to build a book recommendation system that meets the needs of readers and allows them to enrich their reading and satisfaction.

\newpage
The article is organized in the following manner. First, in section \ref{section2}, the work related to the topic of this study is reviewed. Then, in section \ref{section3}, the research design and the method used are described. Next, the results of the experiment are presented in section  \ref{section4}. Finally, in section  \ref{section5}, the results are discussed and section  \ref{section6} concludes by describing the general contributions of the research as well as future directions.

\section{Related works}
\label{section2}
This study proposes a framework to support decision making regarding book recommendation, which integrates techniques such as NLP and CBR. Previous works on relevant aspects are reviewed in the following parts.

\subsection{CBR origin and evolution}
The first contributions in the area of case-based reasoning (CBR) were from \cite{schank1977p}, where the use of CBR as a conceivable high-level model for cognitive processing was highly appreciated. Later, in 1994, \cite{aamodt1994case} successfully used it in several domains, such as prediction, control, and planning. With the evolution of Artificial Intelligence (AI), many works focus on incorporating this type of reasoning into AI systems: \cite{schank1983dynamic, resnick1997recommender, yu2003multi, wang2012recommender, wang2020emergency} among other studies that have been developed to date present. Recently a study by \cite{adelomou2022quantum} has linked case-based reasoning with the quantum world, where more complex algorithms combined with quantum are used to determine similar cases in the Recovery phase.
If the review focuses on the publishing sector, the study carried out by \cite{chang2005hybrid} even offers us the possibility of merging Artificial Intelligence algorithms within the same system to improve its accuracy. Research by \cite{chen2008herd} shows that customers are more interested in books labeled "customers who bought this book have also bought" than books labeled "recommended by bookstore staff" . In this case, it combines it with unsupervised machine learning techniques such as self-organizing maps (SOMs) to forecast book returns to publishers. Later \cite{chang2006hybrid} offers the same approach, this time achieving a hybrid between genetic algorithms (GA) and CBR.
From the literature reviewed, it is observed that this type of techniques are not sufficiently exploited in the publishing field, especially in the recommendation of books that meet the needs of the reader, based on the writing, since the few that exist usually deal with this type of problem. with a global aspect such as the most frequent sales and purchases.

\subsection{Combination of techniques: CBR using NLP}
As the work focuses on recommendation systems in the publishing field, specifically based on the text of books, the literature is searched for studies referring to the combination of CBR techniques using NLP. Only one relevant study has been found, although it was outside the editorial domain. At least the authors of this research have not found related works to date. The first study by \cite{wu2020ontological}, allows the recovery of metro accident cases through the development of an ontological model using the NLP technique for decision making, given that the memory cases consisted of historical records of accidents. This allows, although not from the same domain, to get a little closer to the idea of this research and delve into the similarity of the texts to provide similar cases of a CBR.

\subsection{Similarity of texts in Spanish through NLP}
\begin{figure}[ht!]
\centering
\includegraphics[width=0.80\columnwidth]{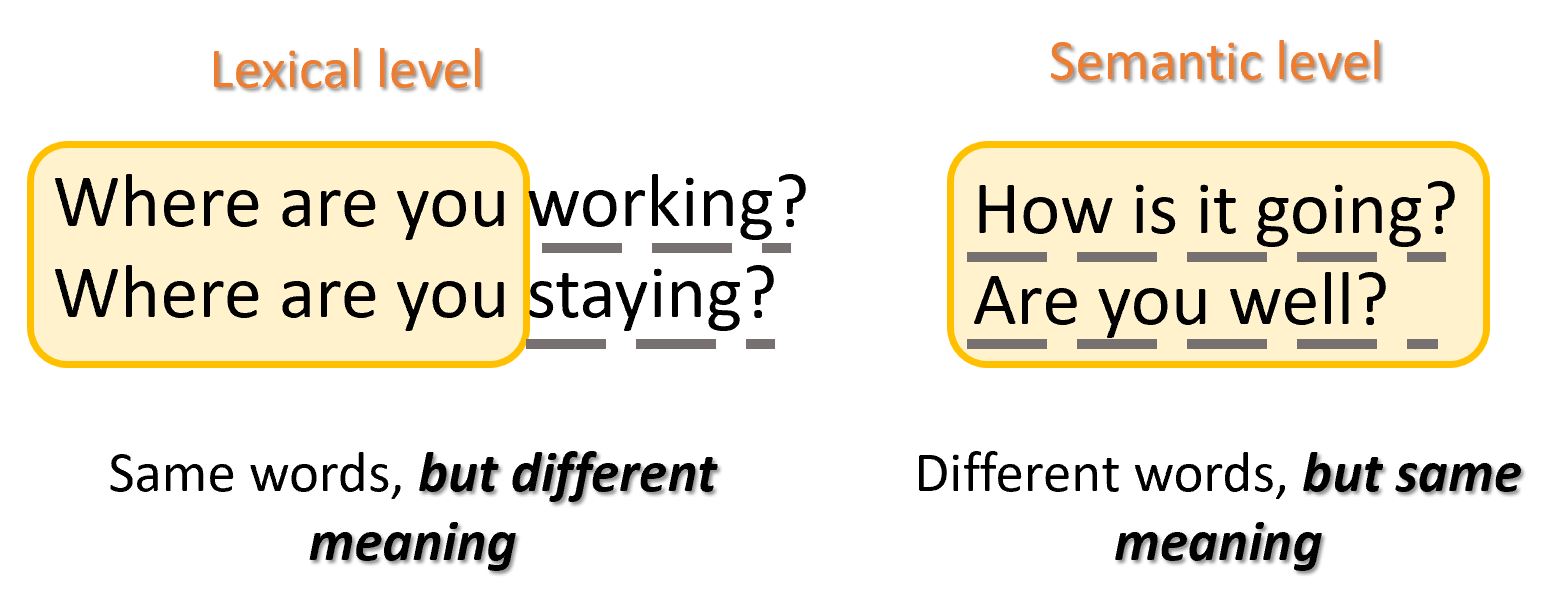}
\caption{Lexical analysis versus semantic analysis in texts. Lexical analysis is based on similar words even though the meaning is different. While the semantic analysis, although it has different words, they present the same meaning. Source: Own elaboration.}
\label{fig1}
\end{figure}

For the recommendation of books, based on the writing, it is important to analyze the text both at the lexical level and at the semantic level. The use of one or the other will depend on the task to be carried out, for example at the lexical level the words will be organized in groups or fields of meaning and at the semantic level they will be associated because they belong to the same grammatical category and share a part of their meaning. As an example, Figure \ref{fig1} shows the difference of both levels.

\begin{figure}[ht!]
\centering
\includegraphics[width=0.80\columnwidth]{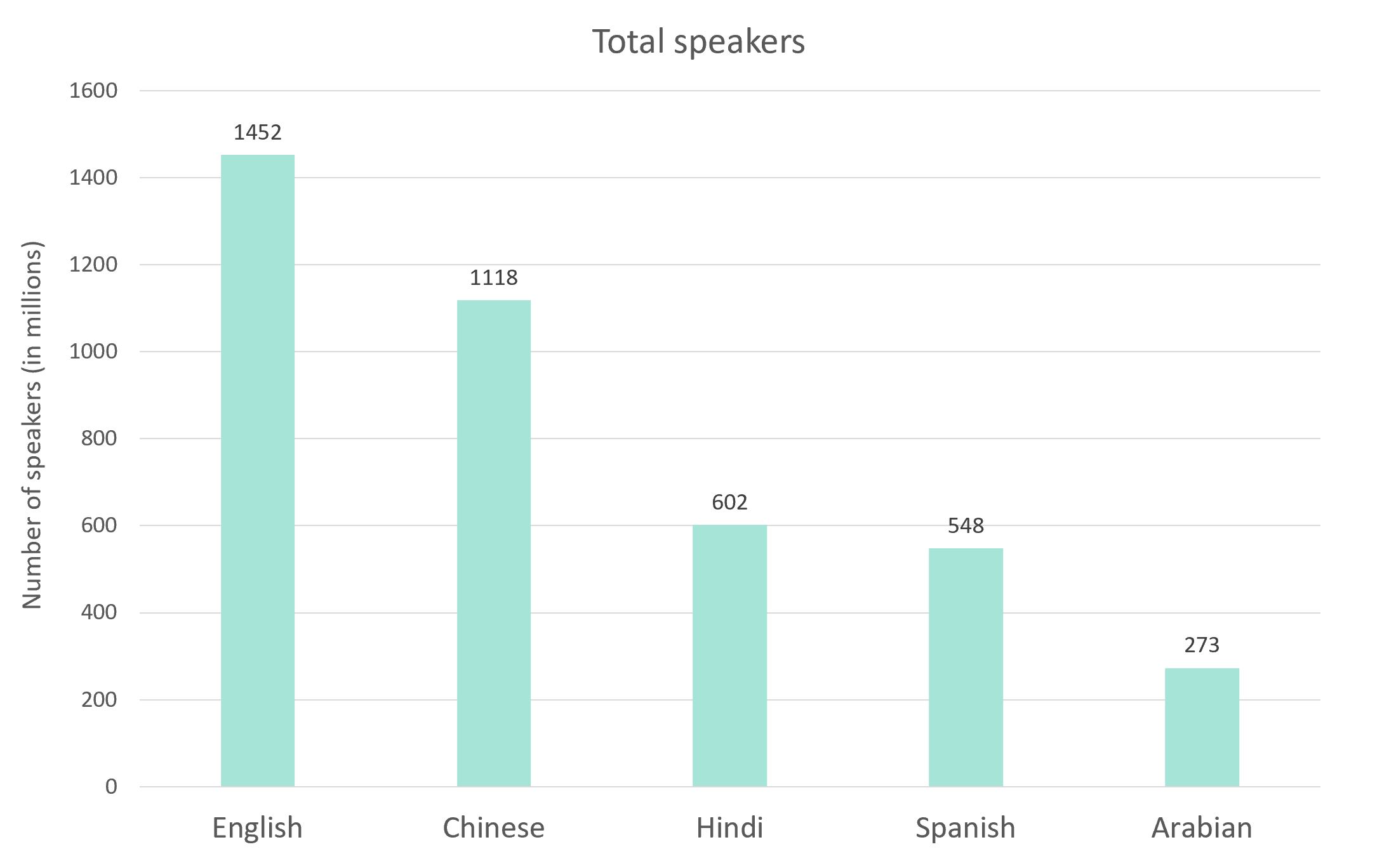}
\caption{The most spoken languages in the world, with Spanish being the 4th most spoken. Source: Own elaboration based on \url{https://es.statista.com/estadisticas/635631/los-idiomas-mas-hablados-en-el-mundo/}}
\label{fig2}
\end{figure}

\newpage
However, the vast majority of studies, such as \cite{salager1991text, gomaa2013survey} on text similarity have focused mainly on the English language. It which means that there are not as many resources as would be needed when working with the Spanish language. Of the few studies found, there is a work of \cite{lopez2016aproximacionn}, where they use the vector representation based on word embeddings for the task of semantic similarity of texts, using evaluation metrics such as Euclidean distance and cosine. Finally, with all these studies, a base is taken of the evaluation metrics and the different approaches that you can take when analyzing a text; and the scarce resources that still exist in the Spanish language are reaffirmed, despite being the 4th most spoken language worldwide, as can be seen in Figure \ref{fig2}.

\newpage
To summarize the authors to date, they do not find related works on a recommender system in the editorial field based on the similarity of wording. So it leads to the development of the system proposed in later sections.

\section{Research design and method}
\label{section3}
\begin{figure}[ht!]
\centering
\includegraphics[width=1.00\columnwidth]{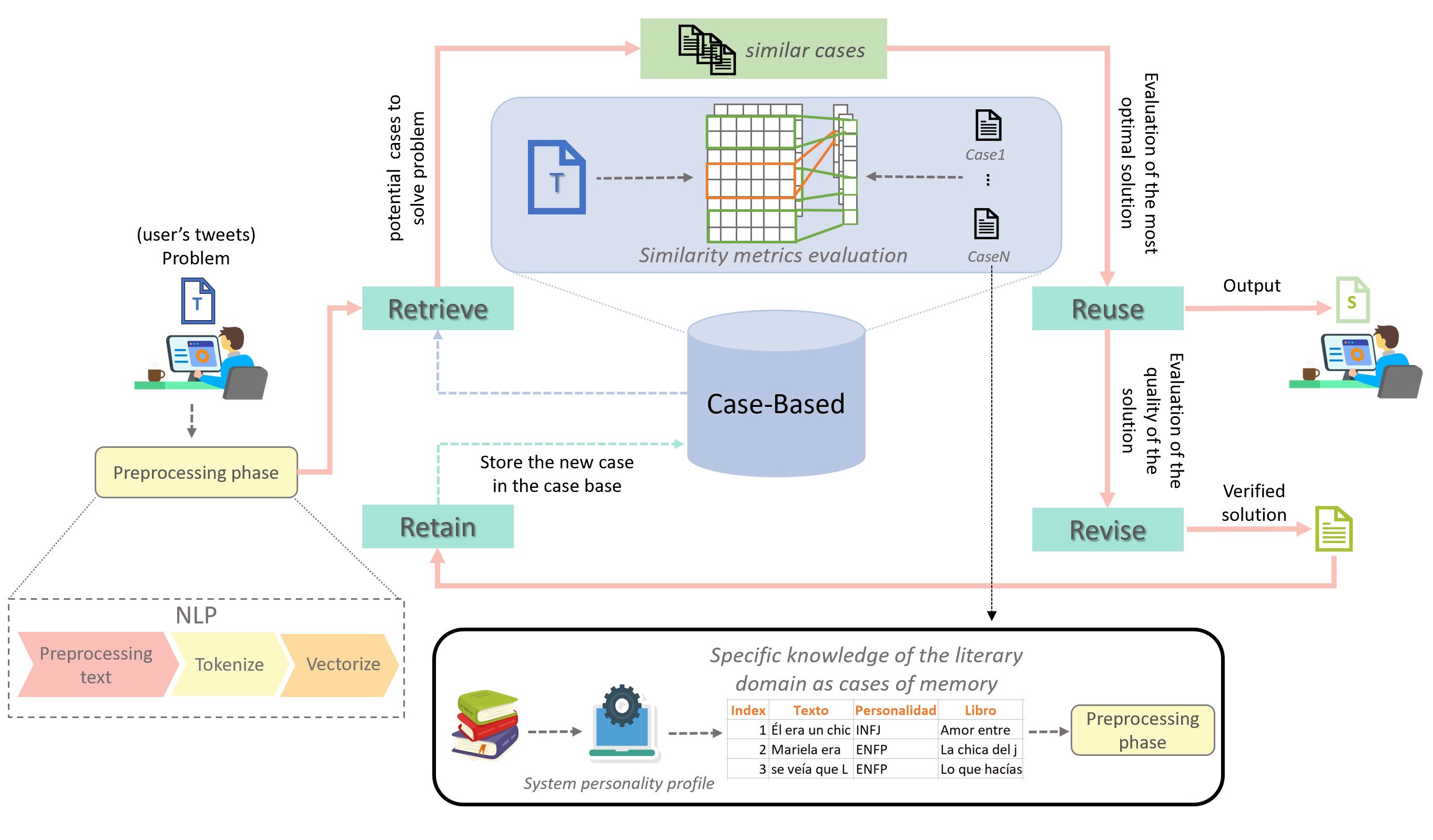}
\caption{Case-based system architecture combined with NLP. Based on how readers express themselves, similar cases are searched in the database. For this, NLP techniques are used for text processing and similarity evaluation metrics between them. Finally, the best case is returned. This case can be reused if an expert validates it and considers it appropriate to retain in the case memory. Source: Own elaboration.}
\label{fig3}
\end{figure}

This section describes the system called, Artificial Intelligence Based Reasoning with Natural Language (BRAIN L), for recommending books based on readers' own writing. The general design of the system is shown in Figure \ref{fig3}.

\newpage
In the following subsections, each of the phases of the system will be explained in detail.
\subsection{Data}
\begin{figure}[ht!]
\centering
\includegraphics[width=1.00\columnwidth]{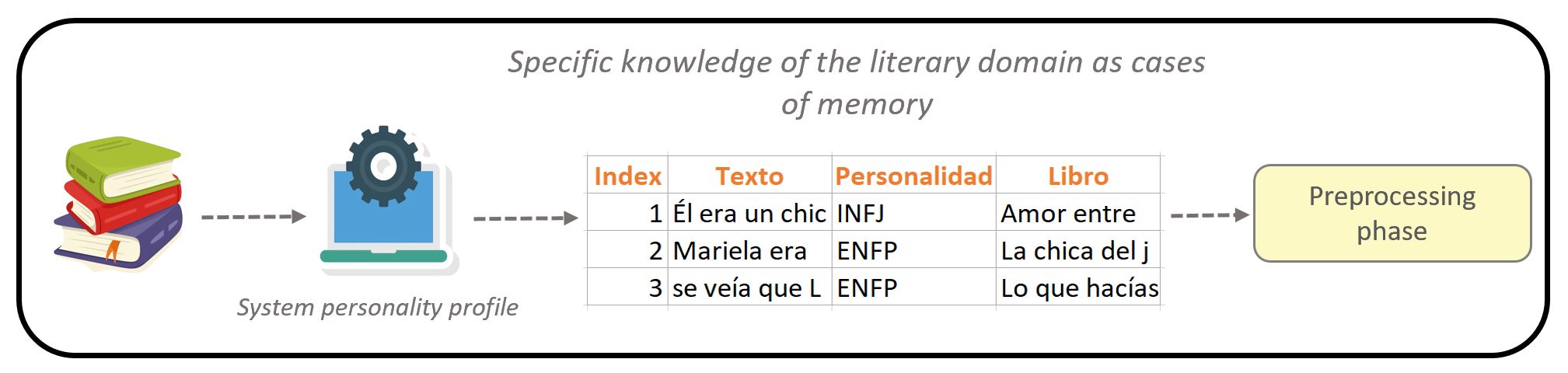}
\caption{Data stored within the memory of cases, resulting from a previous system for determining the personality of the characters. Source: Own elaboration.}
\label{fig3.1}
\end{figure}

The memory cases are based on data obtained as the output of a system for determining the personality of the characters created by \cite{martin2022personality}. A more detailed view of the content is seen in Figure \ref{fig3.1}. Containing the text belonging to each character in the book, the title of the book it belongs to, and the personality associated with the character according to the Myers-Briggs indicators. It has around 150 initial cases, although this database will increase as the experts validate it. 

\newpage
To store these cases, make it easier for them to be shared from different readers at the same time, avoid redundancy and improve the organization of our system, a MongoDB database has been used.

To obtain the readers' way of expressing themselves, data has been collected from the social network Twitter, the second social network with the highest number of users and the easiest to access its data through its own API. Data have been saved in a MongoDB database. It has around 2,590,705 samples from the period from April to March 2022.

\subsection{Retrieve phase}
During this phase, the texts expressed by the reader (Twitter user) are received as the problem to be solved (what books do you recommend based on my expressions?). It is important to declare that both for the texts of the reader, as well as for the cases stored in the case base, they are previously preprocessed, under the same 3 steps that are observed in Figure \ref{fig3}: Text preprocessing, Tokenization, Vectorization. Once both texts are vectorized, the similarity between them is determined.

\newpage
There are different metrics for evaluating the similarity of texts depending on the grouping of the texts and the embeddings performed. The metrics to use are:
\begin{itemize}
    \item Cosine: It is a measure of the similarity between two vectors in a space that has a product in its interior with which the value of the cosine of the angle included between them is evaluated. Figure \ref{fig4} shows the calculation of the metric for a better understanding of the concept. This calculation uses Eq. \ref{eq:eq1}.

    \begin{figure}[ht!]
    \centering
    \includegraphics[width=0.40\columnwidth]{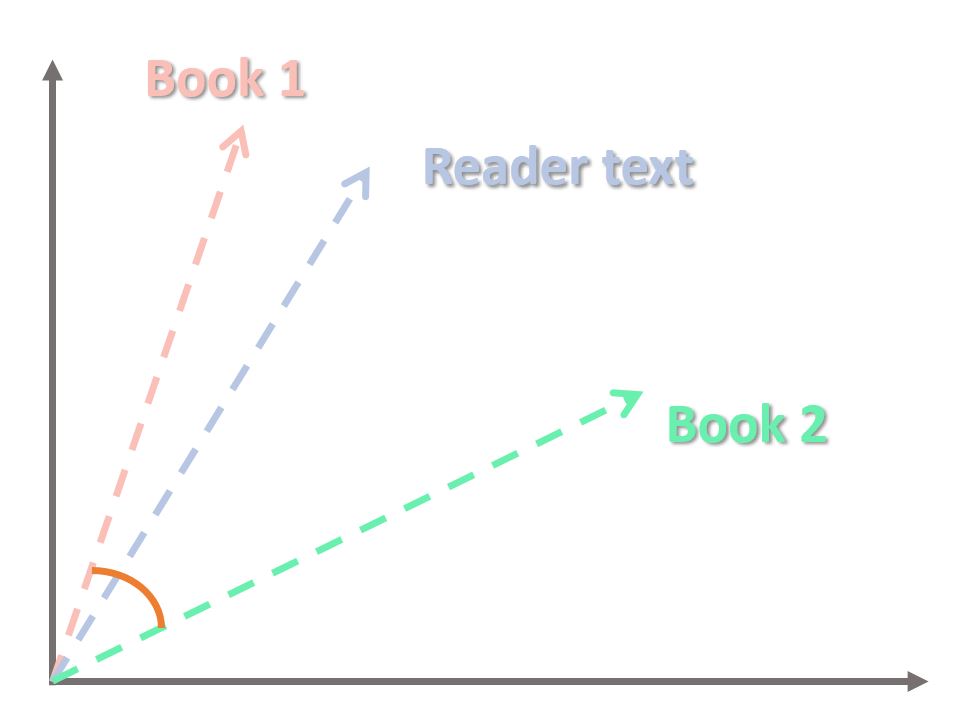}
    \caption{Calculation of cosine similarities between two books and a text entered by the reader (Twitter's user). Source: Own elaboration.}
    \label{fig4}
    \end{figure}

    \begin{equation}\label{eq:eq1}
    Cosine (\theta)=\frac{A \cdot B}{\|A\|\|B\|}=\frac{\sum_{i=1}^{n}A_{i}B_{i}}{\sqrt{\sum_{i=1}^{n}A_{i}^2}\sqrt{\sum_{i=1}^{n}B_{i}^2}}
    \end{equation} 

    \hspace{-2.0em} where \textit{A} and \textit{B} are the texts to compare.

    \item Jaccard: The Jaccard index measures the degree of similarity between two sets, regardless of the types of elements, since it is based on the intersection of the sets. Figure \ref{fig5} shows the calculation of the metric for a better understanding of the concept. For the calculation, Eq \ref{eq:eq2} is used.
    
    \newpage
    \begin{figure}[ht!]
    \centering
    \includegraphics[width=0.40\columnwidth]{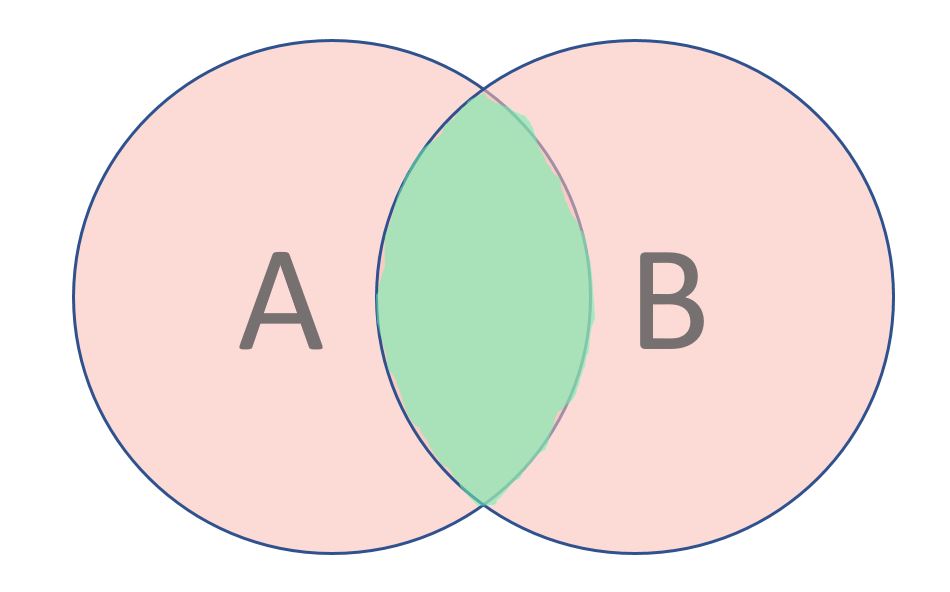}
    \caption{Venn diagram of the two texts to be analyzed. Where A is the text entered by the reader and B the text to be compared from the book. And the similarity is calculated by dividing the intersection (green color) and the union of the texts (pink color). Source: Own elaboration.}
    \label{fig5}
    \end{figure}

    \begin{equation}\label{eq:eq2}
    Jaccard (A,B)=\frac{\|A \cap B\|}{\|A \cup B\|}
    \end{equation} 
    \hspace{-2.0em} where \textit{A} and \textit{B} are the texts to compare.    

     \item SoftCosine: This measure of similarity starts from the same basis as cosine similarity, but generalizes it further by considering similarities between pairs of features \cite{sidorov2014soft}. This calculation uses Eq \ref{eq:eq3}.

    \begin{equation}\label{eq:eq3}
    Softcosine(a,b)=\frac{\sum_{i,j}^{N}s_{ij}a_{i}b_{j}}{\sqrt{\sum_{i,j}^{N}s_{ij}a_{i}a_{j}}\sqrt{\sum_{i,j}^{N}s_{ij}b_{i}b_{j}}}
    \end{equation} 

    \hspace{-2.0em} where the matrix \textit{$s_{ij}$} represents the similarity between features.
    
\end{itemize}

Regarding the embeddings these are:
\begin{itemize}
    \item Word2Vec: Embeddings are performed using a neural network model to learn word associations from a large corpus of text. For this reason it is good for semantic analysis. \cite{mikolov2013linguistic}
    \item GloVe: It is a type of word embedding that encodes the co-occurrence probability relationship between two words as vector differences. \cite{pennington2014glove}
    \newpage
    \item BERT: It is the most advanced technique in the NLP world. It allows extracting characteristics of the embeddings made to the text, useful for semantic information. And furthermore, it adds a special MASK token that allows it to make predictions, finding the words that provide relevant information to the text. \cite{devlin2018bert}
\end{itemize}

Once the potential cases to solve the problem have been determined, in this case, the most similar books according to the reader's writing, the next phase is continued.

\subsection{Reuse phase}
This phase performs an evaluation of the most optimal solution and returns it as the proposed solution for the new problem. It is likely that the retrieved case, if similar enough, is likely to contain a suitable solution. To do this, the threshold is set at 50\%. This allows that once the similar cases have been recovered and given that they present a score; it is analyzed whether or not the highest value exceeds this threshold. In cases where it is not passed, 2 possible books are recommended (from highest to lowest score) with a recommendation reliability message "Recommendation reliability: -50\%". While if it passes it, the case with the best score is displayed and the message "Reliability of the recommendation: +50\%". If the case exceeds the threshold, it indicates that it is apt to be stored as a possible case within the case base, thus moving on to the next phase.

\subsection{Revise phase}
A CBR agent usually requires some feedback to know what is going well and what is going wrong. It is usually done by simulation or by asking a human oracle. For this reason, during this phase, policies are established in the system so that experts in the field can validate whether or not the new case is appropriate to be included in the case base.

\newpage
The policies to be established:
\begin{itemize}
    \item If the case is validated by a single expert, the case will not be inserted.
    \item If the case is validated by two experts, the criterion of not inserting the case must be justified.
    \item If the case is validated by three experts, it will be inserted directly into the database.
\end{itemize}

\subsection{Retain phase}
If the case solution generated during the review phase needs to be retained for future problem solving, the case base is updated with a new case learned in the retention phase. The data that is stored is: the text of the tweet, the recommended book based on the wording of the tweet and the personality associated with the recommended book. In this particular case, there is no implication in storing this data, since it contains the same fields. \\

Once the design of the system has been understood, and for a better understanding of the system, the behavior of the entire CBR cycle using NLP will be illustrated in the next section with a use case.

\section{Case study}
\begin{figure}[ht!]
    \centering
    \includegraphics[width=1.00\columnwidth]{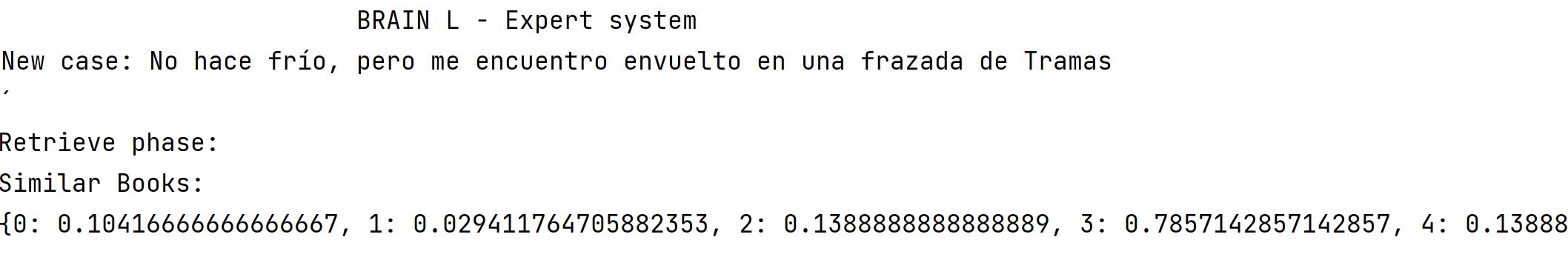}
    \caption{Retrieve phase. The similar cases found in the case base are shown, it contains the index of the book that is similar and the score of the similarity calculation. Source: Own elaboration.}
    \label{fig6}
\end{figure}

\begin{figure}[ht!]
    \centering
    \includegraphics[width=1.00\columnwidth]{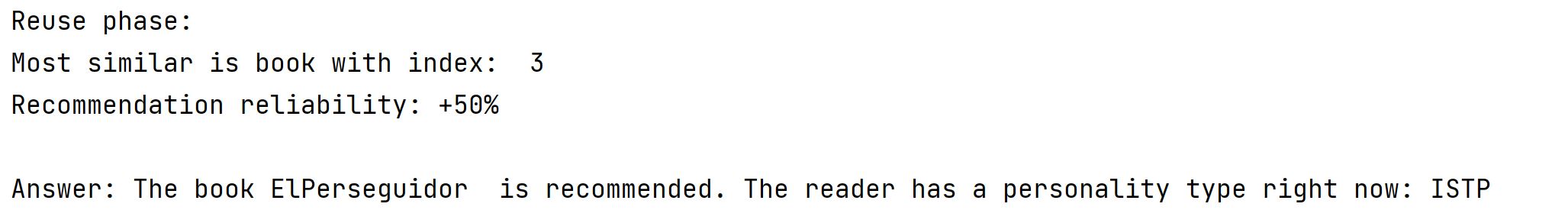}
    \caption{Reuse phase. Obtaining the highest score. It is analyzed if it exceeds 50\% reliability that the recommendation is correct. If you exceed it, that is the case. The reader is shown the recommended book, including the currently associated personality type based on its wording. Source: Own elaboration.}
    \label{fig7}
\end{figure}

\begin{figure}[ht!]
    \centering
    \includegraphics[width=1.00\columnwidth]{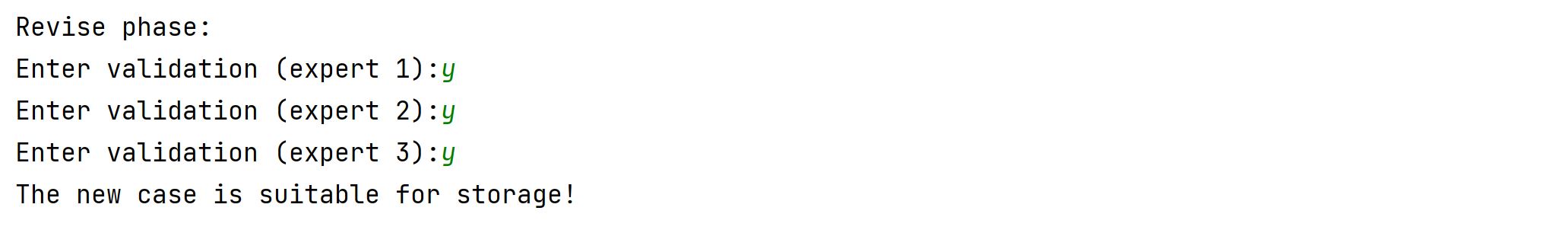}
    \caption{Revise phase. During this phase, the policies are applied so that the experts with the validated solution determine whether or not it should be stored in the memory cases. Source: Own elaboration.}
    \label{fig8}
\end{figure}

\begin{figure}[ht!]
    \centering
    \includegraphics[width=1.00\columnwidth]{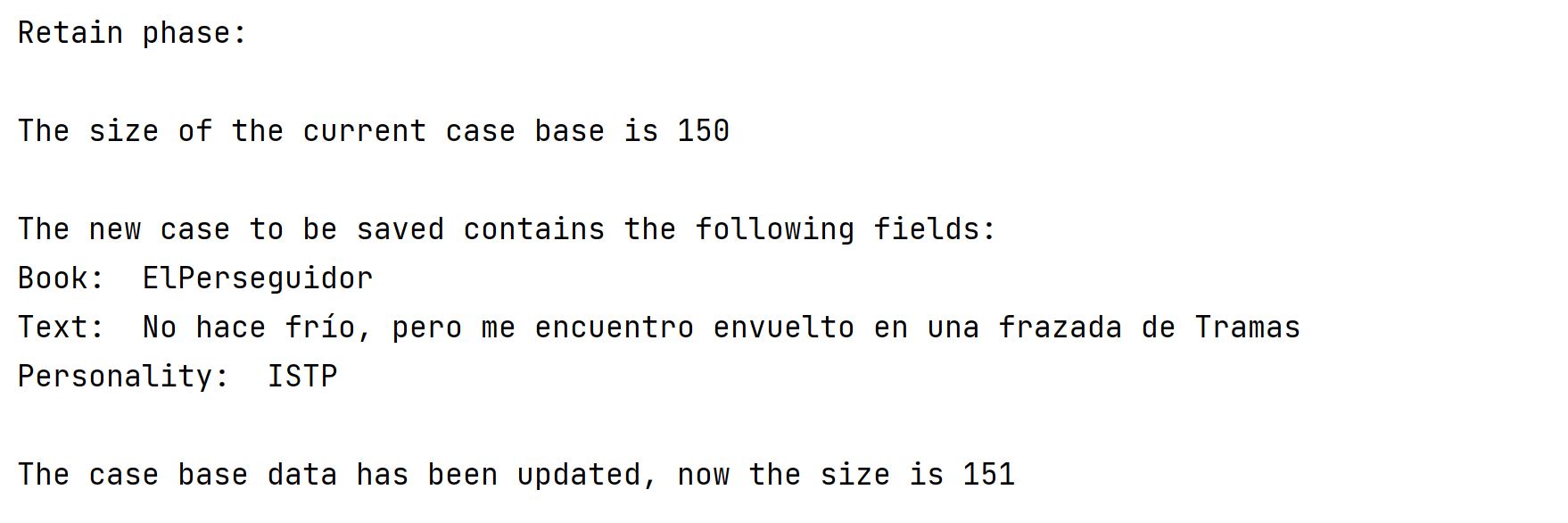}
    \caption{Retain phase. During this phase, the new validated case is stored. Source: Own elaboration.}
    \label{fig9}
\end{figure}

Before showing the results and analyzing them, a case study is proposed, where the complete operation of the CBR is visualized with a real case. This helps to understand the experiments in the following section. Figures \ref{fig6}, \ref{fig7}, \ref{fig8} and \ref{fig9} show the internal flow of each of the phases of the proposed system.

\newpage
The previous image shows how a new case is introduced, specifically, a tweet. When searching for similar cases, internally the system will return the most similar books and the similarity score.

Having several similar cases, internally the system from a threshold established in this case 0.50, assesses which of the scores is the one that exceeds said threshold. This will create an internal validation of the predicted recommendation, before being shown to the end user.

\newpage
If the case selected as optimal and shown to the user can serve as an experience for future searches in the system, certain policies are established. The same if they are shown on the screen to the experts, for their subsequent approval of the insertion of the new case in the memory base.

This case study complies with all the previous phases, and after the unanimous approval of the experts, it is stored within the case base.

\newpage
\section{Experimentation and results}
Once the operation of the system with the case study has been visualized, the results of the experiments carried out are presented.

\label{section4}
\subsection{Data}
The cases stored in the case base come from an older \cite{martin2022personality} system, which returns the character's text, the personality type associated with that text, and the book it belongs to. These data are already preprocessed, in this system they are only tokenized and vectorized in order to perform the similarity calculation.

For the data coming from the readers (on Twitter), if they are preprocessed, eliminating the missing values, since they do not provide any information. As this research is carried out in the Spanish literary domain, the data is limited only to the national territory of Spain. Duplicate data is removed. Only those tweets with more than 20 words are used, thus ensuring that there is valid content for further analysis. Finally it is grouped by reader.

\subsection{Retrieve phase}
For the system training phase, fragments of untrained books are taken as test data, in order to verify the effectiveness of the aforementioned techniques. The modification of the words is also tested, but maintaining the same meaning of the sentence to test the semantic analysis of the techniques. This is the best way to determine if the system will be able to recommend books correctly.\\
During this phase, the different configurations mentioned in the previous section are tested to find the most optimal for our recommendation system. It is important to mention that cosine similarity calculations require embedding in the text; while in Jaccard it is not necessary since it works with the number of elements per game. The configurations made for the calculation of the similarity with Cosine are described below.

For the \textbf{Word2Vec embedding}, the embedding of the spaCy library by \cite{spacy2} itself is used with an \texttt{es core news lg model} and the Cosine similarity is calculated. The results are shown in Table \ref{tab:Table1}.

\begin{table}[h]
    \centering
    \begin{tabular}{c|c|c}
         \hline
            \textbf{Type of text} & \textbf{Similarity} & \textbf{Time(Min)}\\ \hline
            Same text  & 0.99 &  0.25\\
            Different text  & 0.99 &  0.26\\\hline
    \end{tabular}
    \caption{Cosine Similarity Calculation results with Word2Vec Embedding.}
    \label{tab:Table1}
\end{table}

For the \textbf{BERT embedding} the transformer library is used and the data is only worked with the input, which is the part that contains the tokenization of the text, and the attention mask is discarded since no predictions will be made. Since the texts do not have the same length, initially the distance of the longest text is determined and a padding is performed on the shortest text. The copy of the collapsed matrix to a single dimension is also made, so that it has the same shape and the Cosine similarity is calculated. The results are shown in Table \ref{tab:Table2}.

\begin{table}[h]
    \centering
    \begin{tabular}{c|c|c}
         \hline
            \textbf{Type of text} & \textbf{Similarity} & \textbf{Time(Min)}\\ \hline
            Same text  & 0.48 &  204\\
            Different text  & 0.32 &  204\\\hline
    \end{tabular}
    \caption{Cosine Similarity Calculation results with BERT Embedding.}
    \label{tab:Table2}
\end{table}

For the \textbf{GloVe embedding}, the gensim library is used with a \texttt{glove wiki gigaword 50 model} which has the functionality of creating the similarity matrix. And the similarity of SoftCosine is calculated. The results are shown in Table \ref{tab:Table3}.

\begin{table}[h]
    \centering
    \begin{tabular}{c|c|c}
         \hline
            \textbf{Type of text} & \textbf{Similarity} & \textbf{Time(Min)}\\ \hline
            Same text  & 0.91 &  840\\
            Different text  & 0.86 &  860\\\hline
    \end{tabular}
    \caption{Soft Cosine Similarity Calculation results with GloVe Embedding.}
    \label{tab:Table3}
\end{table}

For Jaccard use no embedding is used. The results are shown in Table \ref{tab:Table4}.

\begin{table}[h]
    \centering
    \begin{tabular}{c|c|c}
         \hline
            \textbf{Type of text} & \textbf{Similarity} & \textbf{Time(Min)}\\ \hline
            Same text  & 1.0 &  0.0\\
            Different text  & 0.99 &  0.0\\\hline
    \end{tabular}
    \caption{Jaccard Similarity Calculation results.}
    \label{tab:Table4}
\end{table}

Among the best ones, the similarity calculated with simpler techniques such as Jaccard and Word2Vec can be highlighted. The difference between one technique or another is hardly noticeable, but in a matter of seconds of execution, it could be decided to use Jaccard.

\subsection{Reuse phase}
\begin{figure}[ht!]
    \centering
    \includegraphics[width=1.00\columnwidth]{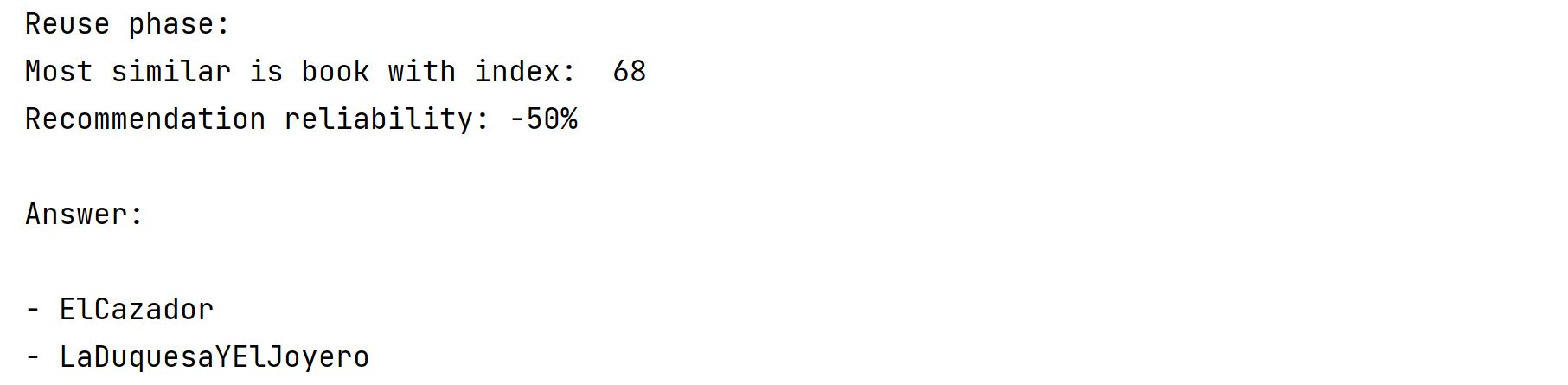}
    \caption{In case the established threshold is not exceeded, an explanation message is displayed, although the two books with the best score are offered even if the threshold is not exceeded.. Source: Own elaboration.}
    \label{fig10}
\end{figure}

During this phase, it helps to increase the validity of similar cases retrieved. Firstly, the recovered cases are ordered from highest to lowest score. The best score is compared with the proposed threshold (50). If it exceeds it, it is shown to the user, otherwise a message is shown stating that the reliability of the recommendation is less than 50\%.\\
For better compression, Figure \ref{fig10} shows another case study where the threshold is not exceeded. For this, the following tweet is used: ``Sonreír es lo más saludable que puedes hacer a diario''.

\newpage
\subsection{Revise phase}
During this phase the experts will determine whether or not the new case should be entered into the case base. For this, functions are implemented that allow validating the policies mentioned in the previous section. The operation of each of the implemented policies is shown below, where Figures \ref{fig11} and \ref{fig12} show the rejected cases.

\begin{figure}[ht!]
    \centering
    \includegraphics[width=1.00\columnwidth]{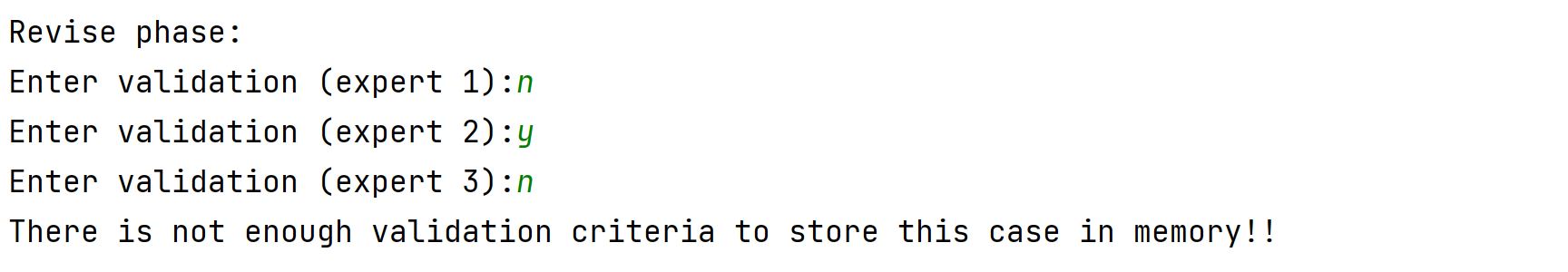}
    \caption{First policy applied, if a single user is the one who thinks that it should be inserted, it will not be taken into account. Source: Own elaboration.}
    \label{fig11}
\end{figure}

\begin{figure}[ht!]
    \centering
    \includegraphics[width=1.00\columnwidth]{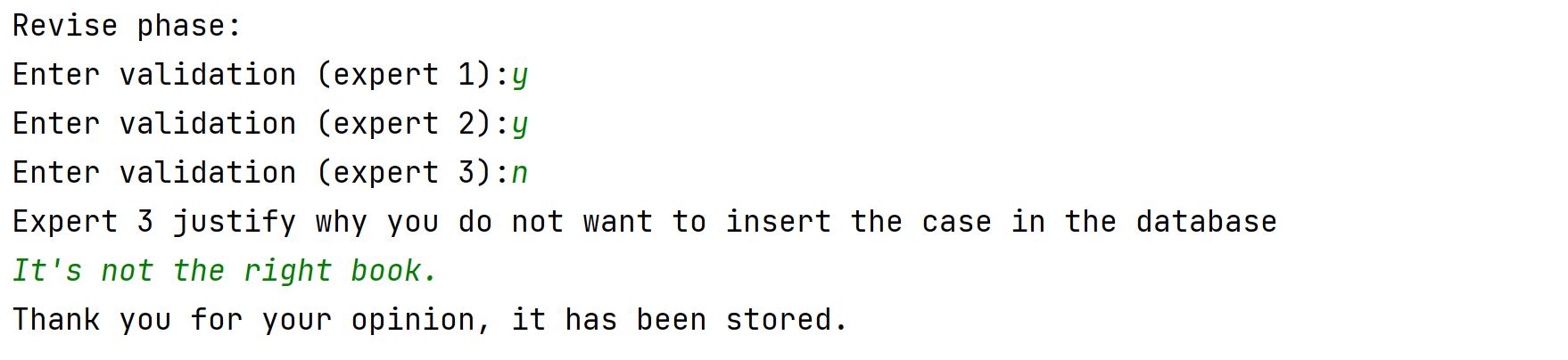}
    \caption{Second policy applied, if a single user is against it, they must store the justification of why the case should not be saved. Source: Own elaboration.}
    \label{fig12}
\end{figure}

\begin{figure}[ht!]
    \centering
    \includegraphics[width=1.00\columnwidth]{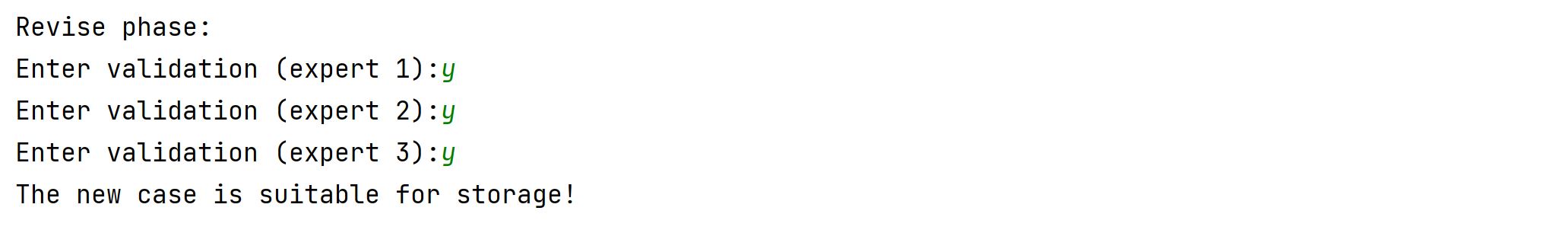}
    \caption{Third policy applied, all experts agree that it should be inserted. Source: Own elaboration.}
    \label{fig13}
\end{figure}

The third policy is shown in Figure \ref{fig13}, where once the experts unanimously validate that the new case is apt to be saved in the case base, the next phase is continued.

\newpage
\subsection{Retain phase}
\begin{figure}[ht!]
    \centering
    \includegraphics[width=0.80\columnwidth]{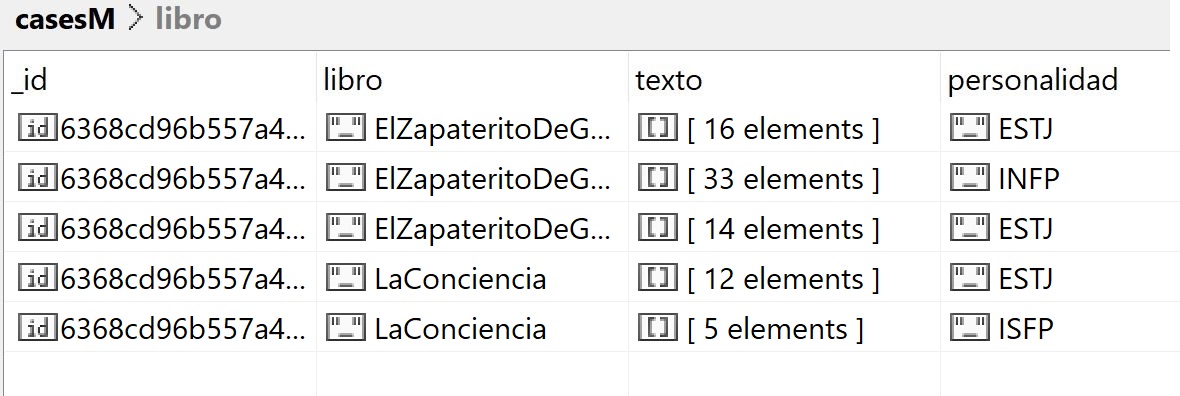}
    \caption{Base of cases with 150 samples, in this image only the last 5 are displayed. Source: Own elaboration.}
    \label{fig14}
\end{figure}

\begin{figure}[ht!]
    \centering
    \includegraphics[width=0.80\columnwidth]{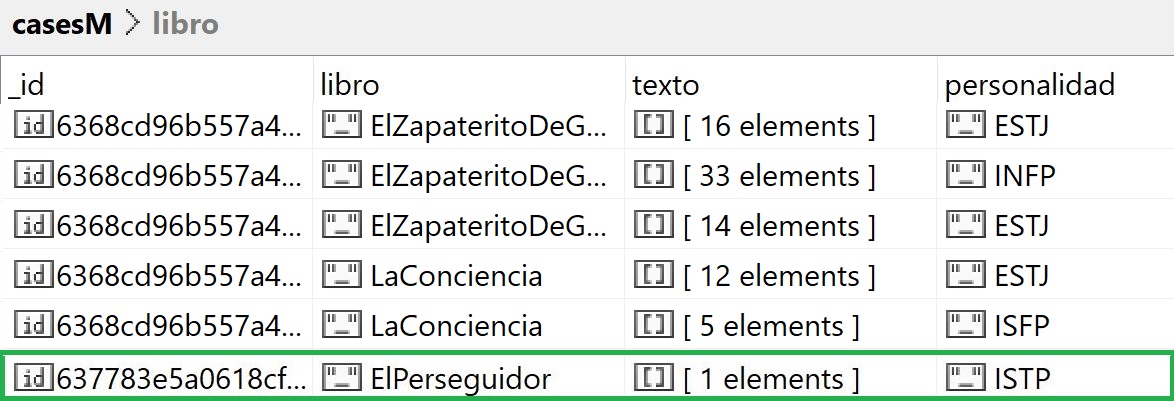}
    \caption{Case base updated with the new case inserted. Source: Own elaboration.}
    \label{fig15}
\end{figure}

To better understand the data that is stored in our case memory, Figure \ref{fig14} shows how the case base is found, with the initial data stored and in Figure \ref{fig15} how it is found after storing the new case, once validated by the experts.

\newpage
\section{Discussion}
\label{section5}
In previous sections, the process of creating the BRAIN L system has been presented, which is a little more in line with what the reader transmits than with historical sales. For this we have based ourselves on the idea of \cite{wu2020ontological}, since of all the related works reviewed to date it is the one that most resembles the idea that we wanted to work on in this research.\\
However, his idea was based more on ontologies than on the text itself, and it is the main contribution that we make with our work. Based on the semantics of the texts, the aim is to recommend the book that best suits the reader, and thereby also determine the psychological profile of the reader, thanks to previous work done by \cite{martin2022personality}.\\
It has been shown during the experiments carried out that although the most advanced embedding technique is BERT, it is not yet ready for calculating text similarity. The resulting data is unreliable and has a high runtime cost. It is shown that with simple techniques in NLP such as Jaccard or Word2Vec, good results can be obtained, reaching up to 99\% accuracy in predictions. The final selection for the Jaccard technique was based on the execution time, which by seconds is much more optimal than Word2Vec.\\
Obviously we are aware that memory cases must be expanded and that we must have the vital experience of experts to validate the new cases entered into our system, but with this work we want to sow a seed of how to improve recommendation systems in the literary sector, where it is currently only recommended based on historical sales.

\newpage
\section{Conclusions and futher work}
\label{section6}
In conclusion, despite the initial data limitations in the BRAIN L system, promising results can be observed, reaching 99\% similarity between texts with different words and the same meaning. The authors of this work, to date, have not found a complete CBR system that focuses on the publishing sector and especially on the recommendation of books focused on the text itself and not on historical sales. Keep in mind that the recommended books  with the reader can be variable, because they are based on how the reader writes at a given time on Twitter, which gives a real validity to the system, since it recommends based on how we express ourselves in different moments of the day. With the promising results obtained from the experiments carried out, we will contribute to stimulating the reader's empathy with the character, and therefore, to encourage reading since it will adapt more to the reader's personality. It is necessary to emphasize the fact that the system is in Spanish. This provides an extra contribution to the work and community of NLP in Spanish, which still has a long way to go.

Many different adaptations, tests and experiments have been left for the future due to limitations. Some of them are summarized here: a) The results obtained are limited in terms of recovery of cases within a small database of cases (150 records), the size of the database should be expanded; b)Expand the recommender and take into account authors, genres, among others; c) Put this study into production once it has been validated with sufficient data; d) Increase validation policies and e) Create an interface that is friendlier to the end user.

\section*{Acknowledgements}
It has also been possible thanks to Smart Society research group at La Salle--Ramon Llull University. We would like to give a special thanks to those open source institutions that contribute to the possibility of these types of studies and promote literary culture, to facilitate the personal, cultural, professional and social development of people.

\newpage
\bibliography{sample}

\end{document}